\documentstyle[12pt, fleqn]{article}
\begin{document}
\title{ Inclusive Photoproduction of $\eta$ Mesons on  Nuclei and 
the in-medium  properties of the S$_{11}$ Resonance  
\thanks{ Work supported in part by the Natural Sciences and
            Engineering Research Council of Canada} }
\author{\bf  M. Hedayati-Poor$^{a,b}$ and H.S. Sherif$^a$ \\
(a)Department of Physics, University of Alberta \\
Edmonton,  Alberta, Canada T6G 2J1\\
(b)Department of Physics, University of Arak, Iran}

\date{\today}

\maketitle

\begin{abstract}
A relativistic non-local model for the inclusive
photoproduction of $\eta$
mesons from complex nuclei is introduced. The model is based
on the dominance of the S$_{11}$(1535) resonance. We compare
the results of our calculations with the available data on
inclusive cross sections for the nuclei C, Al and Cu.
Assuming the resonance propagates freely in the nuclear
medium, we find that the calculated angular distribution
and energy dependence of the cross sections 
reproduce the data in a reasonable fashion. The present non-local
model allows the inclusion of density dependent mass and
width in the calculations. Including these in the
calculations reveals that the presently available data do
not show clear preference for the inclusion of such
modifications of the properties of the S$_{11}$(1535) in the
nuclear medium.
\end{abstract}
\hspace{-.25 in}PACS number(s)  25.20.Lj, 24.10.Jv, 14.20.Gk, 13.60.Le

\newpage
\label{intro}
\section{Introduction}
The question of possible modifications of hadron properties in the nuclear
medium has attracted considerable interest in recent years. 
This interest stems
from the hope that studies of these effects can clarify the interplay of the
competing degrees of freedom in the nucleus. In particular it may be possible
to learn about explicit quark degrees of freedom in the nucleus through such
studies.

There have been several theoretical as well as experimental investigations of
these effects, and evidence is gathering that indeed such modifications do
take place. Several studies have been carried out for the modifications 
of meson
properties in heavy ion and  other collisions \cite{BR,PR,UM,STT,DA}. 
Photonuclear reactions also lend
themselves to investigations of this type \cite{CTC, UM, BKE}. 
In particular, a promising area of
investigation is the photoproduction of $\eta$ mesons on nuclei in the 
energy range corresponding to the second resonance region of the nucleon. 
The fact that one
resonance, the S$_{11}$(1535), dominates the production process at these
energies, makes it attractive to look at these reactions for clues on 
changes in
its properties in the nuclear medium. 

Three pioneering and complementary  measurements of the inclusive eta
photoproduction on nuclei have been reported in recent years. The earlier
measurements of Robig-Landau {\it et al} \cite{RLPLB} at MAMI included cross 
sections on a number
of nuclei from threshold up to a photon energy of 800 MeV. These cross sections
displayed the expected rise on the low energy side of
the resonance. Few years later measurements of the same reaction were 
performed at KEK \cite{YORITA}, extending the
energy range to 1GeV. This 
group reported that a broad resonance due to the excitation and decay of
the S$_{11}$ resonance in the nucleus has been observed for the first time. 
In a recent investigation, Kinoshita {\it et al} \cite{kinoshita} 
reported 
on measurements of this reaction on C and Cu targets at LNS, 
for photon energies up to 1.1 GeV. 
These authors indicated the importance of the contributions from other 
nucleon resonances 
in the second resonance region in addition to the dominant S$_{11}$, at 
energies 
higher than 900 MeV. 
The authors of references \cite{YORITA, kinoshita} performed calculations for 
their data 
using an
adaptation of  the Quantum Molecular Dynamics model. Lehr {\it et 
al} \cite{LM}
performed calculations based on the semi-classical BUU transport model.
Another approach was that of Maruyama and Chiba \cite{MC} who carried out 
relativistic
calculations in infinite nuclear matter based on an effective Lagrangian and a
scaling factor to account for the final state interactions of the $\eta$ 
mesons.

In the present paper we present an analysis of the data based on a
relativistic model for the inclusive photoproduction on finite nuclei. The
 main ingredients of model are the use of an effective Lagrangian
approach (Benmerrouche {\it et al}\cite{ben}) and the relativistic 
mean field
approach to nuclear dynamics \cite{bsjw}. The final state 
interactions of
the outgoing $\eta$ meson are taken into account. This fully non-local model 
has
been discussed recently for the cases of exclusive and incoherent
photoproduction \cite{moh1,moh2}.  These non-local calculations, in contrast 
to earlier 
local ones \cite{moh3}, allow us to include possible density dependent 
changes in the mass 
and width of 
the resonances in the nuclear medium. 

The justification for the claim that the inclusive data
from KEK, Mainz and LNS 
may reveal changes in the properties of the S$_{11}$ resonance requires some
discussion. It is known that the S$_{11}$ is the main contributor to the 
$\eta$ photoproduction 
cross
section on free nucleons in the 0.6 - 1GeV range. There are contributions from
neighboring resonances and from nucleon Born terms as well as meson exchange 
diagrams;
but these are small by comparison. We will show in the course of our
discussion that this situation carries over to the inclusive production on
nuclei. 
The contributions are only at the few percent level, a situation
which allows us to probe whether the existing data 
does reveal any information regarding the properties of 
the S$_{11}$ in the medium. 
As we shall show there are remaining uncertainties regarding  other aspects
of the analysis which are more important compared to the small contributions
of other diagrams.

The plan of the paper follows a recent discussion of the 
non-local relativistic model for 
the exclusive quasifree photoproduction of $\eta$ mesons. 
The reaction amplitude
is cast in a general form in section \ref{form} where the method of
evaluation of the amplitude in its non-local form is outlined.  The results 
of the model and the discussion are given in section \ref{results}. 
Conclusions are given in section 
\ref{conclusion}.

\section{Formalism}
\label{form}
In the inclusive reaction, a photon is absorbed by the nucleus and as
a result an $\eta$ meson is produced. This meson is the only
detected particle and there is no information about the final
state of the nuclear system. In the present work it is assumed that 
the inclusive reaction is  based on the 
exclusive reaction in which a single nucleon is ejected in the continum in the course of the production process \cite{bt, moh4}. 
The transition amplitude for the exclusive reaction through the
S$_{11}$ resonance, in its full non-local
form, is given by \cite{moh1},
\begin{eqnarray}
S_{fi} = &&\frac{e}{(2\pi)^{17/2}}\frac{\kappa_R  g_{{\eta} N R}}{
M+M_R} {\left( \frac{M}{E_N} \frac{1}{2\omega_\eta}
\frac{1}{2\omega_\gamma} \right)}^{1/2}\label{eq1}\\
\times&& \sum_{J_B M_B}{\left( J_f,
J_B; M_f, M_B| J_i, M_i \right)}
  \left [S_{J_i J_f} (J_B)\right]^\frac{1}{2}
\nonumber\\
\times&&
 \{\frac{}{} \int d^4 x d^4 y d^4 p \hspace{0.05 in} \bar{\psi}_{sf}(y)
\phi^*_\eta(y)
\frac{ e^{-i p(y-x)}}{ { / \hspace{-0.095 in}p } - M_R +
i\frac{\Gamma}{2} } \gamma_5 { { / \hspace{-0.105 in} k }_{\gamma}
{ / \hspace{-.09 in} \epsilon } }\hspace{0.05 in}   e^{-i
k_{\gamma} x} \psi_B(x) \nonumber
\\
&&+\int d^4 x d^4 y d^4 p  \hspace{0.05 in} \bar{\psi}_{sf}(y)
\gamma_5 { { / \hspace{-0.105 in} k }_{\gamma} { / \hspace{-0.09
in} \epsilon } }\hspace{0.05 in} e^{-i k_{\gamma}y} \frac{ e^{-i
p(y-x)}}{ { / \hspace{-0.095 in}p } - M_R + i\frac{\Gamma}{2} }
\phi^*_\eta(x)\psi_B(x)\hspace{0.05
 in}\frac{}{}\},
\nonumber
\end{eqnarray}
where $E_N$, $\omega_\eta$ and $\omega_\gamma$ are the energies of the
final nucleon, $\eta$ meson and incident photon, respectively.
$M$ is the nucleon mass, $M_R$ and $\Gamma$ represent the mass and
width of resonance. $  S_{J_i J_f} (J_B) $ and $\left(
J_f, J_B; M_f, M_B| J_i, M_i \right)$ are spectroscopic and
Clebsh-Gordon coefficients. $\kappa_R $ and
$g_{{\eta} N R}$ are the  anomalous magnetic moment of the resonance
and its coupling constant with $\eta$ and nucleon.
$\psi_{sf}(y)$, $\psi_{B}(y)$ and $\phi_\eta(y)$ are wave
functions describing the final nucleon, initial nucleon and
$\eta$ meson, respectively. $\psi_{B}(y)$ is a solution of the
Dirac equation with appropriate mean field potentials and
$\phi_\eta$ is a solution of the Klein - Gordon equation. As is
explained in \cite{moh4}, the wave function of the final nucleon
($\psi_{sf}(y)$) is taken to be a plane wave. Calculations without 
inclusion of the final state interaction of $\eta$ meson
 revealed that the u-channel contribution (second
term in (\ref{eq1})) is less by two orders of magnitude than
that of s-channel (first term in (\ref{eq1})). 
As a result the non-locality
and possible  medium modifications of the former will have 
a negligible effect on
the calculated cross section. We thus ignore this very small change and
treat the u-channel locally as in \cite{moh4}. The dominant s-channel 
diagram is treated in the non-local approach. 

The medium modifications of the properties of the nucleon resonance are often 
expressed  in terms of changes of its mass and width from their free values in 
a manner dependent on the nuclear density \cite{ST,LM}. Our
non-local approach can accommodate these density dependent
parameters for the resonance. After implementing the density
dependent mass and width for the resonance in the s-channel  part
of  the above amplitude, we
 calculate the resulting integral following an approach similar to 
the one used in ref.\cite{cooper}.
We rewrite the  integral of the s-channel part in eq(\ref{eq1}) as
\begin{eqnarray}
(2\pi)^{4} \int d^4 y \hspace{0.05 in}\chi^\dagger_{sf} ( 1\hspace{.1 in}
+\frac{\sigma\cdot {\bf k_N}}{E_N +M}) e^{i k_N
y}\phi^*_\eta(y) W(y), \label{eq2}
\end{eqnarray}
where we use a plane wave function (see ref.  \cite{moh4}) for the final 
state of the struck nucleon; $\chi_{sf}$ denotes its spin state. The function
$W(y)$ has the following form
\begin{eqnarray}
(2\pi)^{4}W(y) &=& \int d^4 x d^4 p \frac{ e^{-i p(y-x)}}{ { /
\hspace{-0.095 in}p } - M_R(\rho) + i\frac{\Gamma(\rho)}{2}}
 \gamma_5
{ { / \hspace{-0.105 in} k }_{\gamma} { / \hspace{-.09 in}
\epsilon } }\hspace{0.05 in}   e^{-i k_{\gamma} x} \psi_B(x).
\label{eq3}
\end{eqnarray}
Acting  with the operator
 ${ / \hspace{-0.095 in}p } - M_R(\rho) + i\frac{\Gamma(\rho)}{2}$ 
($\rho$ refers to the density of the nucleus) from the left on  
(\ref{eq3}) and then
carrying the integration over momentum, the following Dirac-type
linear differential equation is obtained
\begin{eqnarray}\label{eq4}
  ( \hspace{0.05 in}{ / \hspace{-0.095 in}p }-M_R(\rho) 
+ i\frac{\Gamma(\rho)}{2} )W(y) = V(y),
\end{eqnarray}
where the source term is
\begin{eqnarray}\label{eq5}
 V(y) &=& \gamma_5
{ { / \hspace{-0.105 in} k }_{\gamma} { / \hspace{-.09 in}
\epsilon } }\hspace{0.05 in}   e^{-i k_{\gamma} y} \psi_B(y).
\end{eqnarray}
Equation (\ref{eq4}) leads to the following second order
differential equation for the the upper component of $W({\bf r})$.
\begin{equation}\label{eq6}
[{\bf p}^2 - \alpha(r) \beta(r)] W_{up}({\bf r}) =
{\bf \sigma} \cdot{\bf p} V_d({\bf r})-\beta(r) V_{up}({\bf r}),
\end{equation}
where the indices $up$ and $d$ indicate the upper and lower components
of the functions $ W({\bf r}) $ and $ V({\bf r}) $. Note these
functions continue to be spin dependent. The lower component
$W_d({\bf r})$ can be obtained from the upper component as,
\begin{eqnarray}\label{eq71}
  W_d({\bf r}) = \frac{\sigma 
\cdot {\bf p} W_{up}({\bf r}) - V_d({\bf r})}{\beta(r)}.
\end{eqnarray}
The functions $\alpha(r)$ and $\beta(r)$ are
\begin{eqnarray}\label{eq81}
 \alpha(r)& =& E_b + w_\gamma - M_R(\rho) + i\frac{\Gamma(\rho)}{2} \nonumber 
\\
 \beta(r) & =& E_b + w_\gamma + M_R(\rho) - i\frac{\Gamma(\rho)}{2},
\end{eqnarray}
where $E_b $ is the energy of the bound nucleon. 
$W_{up}({\bf r})$ is obtained by solving eq.(\ref{eq6}) following the same  
approach as in ref.\cite{moh1}.

Substituting  the solution of $ W_{up}({\bf r}) $ into eq.(\ref{eq1}) we get,
\begin{eqnarray}\label{eq11}
S_{fi} &=& \frac{ e }{\pi} {\left( \frac{E_p + M}{E_p \omega_\eta
\omega_\gamma} \right)}^{1/2} \frac{\kappa_R  g_{{\eta} N R}}{
M+M_R}\delta( E_B + \omega_\gamma - E_p  - \omega_\eta )
\\
& &\times \sum_{ J_B M_B}{ \left( J_f, J_B; M_f, M_B| J_i, M_i
\right)
                     { \left[ {\cal S}_{J_i J_f} (J_B) \right] }^{1/2}  }
(Z^{s }_{s_f M_B} + Z^{u }_{s_f M_B})
\nonumber 
\end{eqnarray}
The transition function for the u-channel $Z^{u }_{s_f
M_B}$ is calculated in the local approximation and is the same as
eq.(3) in \cite{moh4}. The corresponding function for the s-channel is
calculated in non-local approach described above and is
\begin{eqnarray}
Z^{s }_{s_f M_B} &=&\frac{1}{(4\pi)^\frac{1}{2}}
 \sum_{LJML_\eta,M_\eta}
i^{-(L+L_\eta)}Y^{M-s_f}_L(\hat{\bf k}_f)
\left[Y^{M_\eta}_{L_\eta}(\hat{\bf k}_\eta)\right]^*
\nonumber\\
& &\times\left( L, 1/2 ; M-s_f, s_f| J,M\right)
\\
\times\int &d^3 r &\left[{\cal Y}^{M}_{L,1/2,J}(\Omega)\right]^*
 v_{L_\eta}(r)Y_{L_\eta}^{M_\eta}(\Omega)
( W_u({\bf r}) -\frac{\sigma\cdot {\bf k_f}}{E_N +M}W_d({\bf r}))
\nonumber
\end{eqnarray}
where $v_{l_\eta}(r)$ is the radial wave function of
the $\eta$ meson. $\hat{\bf k}_f$ and $\hat{\bf k}_\eta$ are unit 
vectors along the 
directions of the final nucleon and $\eta$ meson, respectively. The
amplitude (\ref{eq11}) is used to calculate the cross section
for the exclusive reaction (see eq(5) of ref.\cite{moh4}) . 
The inclusive cross section is
obtained by numerical integration of the exclusive cross section
over the phase space of the final nucleon and summing over all energy 
levels of the target nucleus.

In the following section we shall explore whether the available data can 
be used to clarify the issue of possible modifications of the mass and width 
of the S$_{11}$ resonance. For this purpose we will adopt phenomenological 
forms of these parameters suggested by earlier works \cite{ST, LM}.
For mass we use a density dependent form suggested by the quark-meson
coupling model for hadrons \cite{ST}.
\begin{eqnarray}\label{eq7}
M( \rho ) &=& M_{\mbox{free}}(1 - b  \frac{\rho}{\rho^0})
\end{eqnarray}
For width we use  a density dependent form given in \cite{LM} which includes 
the broadening of the resonance width due to its collisions with 
hadrons in the medium,
\begin{eqnarray}\label{eq8}
\Gamma( \rho ) &=& \Gamma_{\mbox{free}} + c \frac{\rho}{\rho^0},
\end{eqnarray}

\section{Results and Discussion}
\label{results}
Throughout the calculations presented here we use scalar and
vector potentials
of Woods Saxon form to generate the bound state Dirac
wave functions of the nucleons.
The parameters of these potentials for different targets are
listed in table
\ref{table1}. The final state interactions of the eta meson
are taken into account through the use of the optical
potential DW1 from ref.\cite{4LEE.NPA}. For the coupling to the 
proton we use the parameters of set 3 given in table 1 of 
ref.\cite{BS} and in calculating 
the neutron contributions we use the factor 
$\frac{2}{3}$ for the ratio of neutron to proton cross sections.

It is well known that the photoproduction process on the
nucleon, in the second resonance region, is dominated by
the S$_{11}$(1535) diagram.  We begin our discussion by
showing that this situation carries over to the
inclusive reactions on nuclei. 
The collective contributions from other diagrams
are small by comparison. This is shown in Fig.1.
Figure 1(a) shows the angular distribution of the inclusive
cross
section of the reaction on $^{12}$C at a photon incident energy
 E$_\gamma$ = 820 MeV.
The solid curve gives the S$_{11}$
contribution while the dashed curve includes the additional
contributions due to the D$_{13}$, the nucleon pole terms
and the
meson exchange diagrams.

The total inclusive cross section on $^{12}$C for photon
energies in the
range 0.65 - 1.1 GeV is depicted in Fig.1(b). 
The dominance of the S$_{11}$
contributions is clearly evident from these comparisons.  It
is worth mentioning
here that this rather fortunate circumstance arises not only
from
the relative smallness of the contributions of the other
diagrams, but
also from strong cancellations among these contributions.

Fig.2  shows the inclusive angular distributions for
C,  Al  and  Cu. For each target nucleus, we
show  the distributions at two photon energies from among
those measured: 740 MeV and 980 MeV. The data are those of
KEK \cite{YORITA}, LNS \cite{kinoshita} and MAINZ (only for
 $^{12}$C at 740 MeV) \cite{RLPLB}.  
The solid curves represent non-local
calculations that assume free propagation for the S$_{11}$
resonance. We take the free mass and width of the resonance
to be 1535 MeV and 160 MeV, respectively. In these and
subsequent figures these curves are labeled 'Free'.
General trends in the data are reproduced and the agreement
is almost quantitative in some cases. The main exception is
the case of Al at 980 MeV.  There are some
disagreements among the measurements from KEK and LNS in the
case of Cu and our theoretical calculations show better
agreement with the latter measurements.

The dashed curves in Fig. 2 represent calculations in which,
following suggestions of ref.\cite{LM}, the width of the
S$_{11}$ resonance is assumed to
undergo density dependent broadening in the medium of the form
given in eq(\ref{eq8}) with c = 50 MeV. We note that the
broadening leads to reduction in the cross sections for all
three nuclei. It may be argued that the effect leads to a
worsening in the accord with the data for C at the lower
energy (740 MeV) and by contrast to a slight improvement at
980 MeV. For Al the quality of agreement with the data
remains about the same, whereas for Cu, the disagreement
between measurements makes it difficult to judge the outcome.

We now consider the effect of a reduction of mass on the
angular distributions. 
In the discussion of ref.\cite{ST}, the
authors suggested that the
nucleon mass in the medium is reduced according to
eq.(\ref{eq7}) with a value of the reduction parameter b = 0.14.
Fig.3 shows the results of the angular distribution
for the same targets and photon energies as in Fig.2.
The dash-dotted curves are obtained when the above value is
used for
the reduction of the mass of the S$_{11}$.  Compared to
calculations
with no medium effects (solid curves), we note a large
decrease in
the cross section. This also brings the calculated cross
section much below the data. This clearly indicates that if
the mass of the S$_{11}$ were reduced in the medium, the
scale of reduction would be much smaller than is
indicated by the value of b given above.
If the reduction
in mass is made weaker the effect becomes smaller. For
example the
dashed curves show
the calculations for a much smaller value of b (b = 0.014).
Here the
effect on the cross section is relatively small. We note
that at the lower energy the effect leads to a slight
increase in the magnitude of the cross section while the
effect at the higher energy is a slight decrease.

The data for the total inclusive cross section (obtained by
integrating the angular distributions discussed above) are
shown in Figs.4-6.
 These include the data for the three nuclei  C,  Al
and Cu from KEK, Mainz and LNS.
Fig.4 displays the inclusive data for C. In Fig.4(a) we
show the effects of changes in the mass
of the S$_{11}$ on the behavior of the total inclusive cross
section in
the energy range from near threshold to 1.1 GeV. The solid curve
represents the case of no medium modification of the mass.
We note that these calculations provide reasonable agreement 
with the
data particularly over and beyond the peak, but are slightly
lower at the lower energies. The dashed and dash-dotted
curves represent the two cases where the mass
modification parameter b of eq.(\ref{eq7}) has the values
0.014 and
0.035, respectively. The latter value, though it improves
the accord with the data
on the lower energy side of the peak somewhat, still gives a
strong suppression on the higher energy side. Calculations
with the smaller value of b lie blew the data at energies
above 900 MeV.

In Fig.4(b) we present similar comparisons involving changes to
the width of the resonance. We show two cases for density
dependent changes in the width of the form given
in eq.(\ref{eq8}). The dash-dotted curve is for a value of c
= 25 MeV
and the dashed curve is for c = 50 MeV. Both these
calculations lead to a lowering of the cross section away
from the data.

Figs.5 shows similar comparisons for the KEK inclusive
cross section on Al and Fig.6 compares the
calculations with the
KEK and LNS data for Cu.
The qualitative features of the results are similar. Again
the data rule out any strong reduction (associated with b =
0.14) in the mass of the resonance. A reduction in
the mass of S$_{11}$ corresponding to a value of b = 0.014
brings the
calculations somewhat closer to the data for Al, but leads
to a reduced cross
Section at higher energies for Cu.  In the
case of the modification of width we note that the density
dependent width of eq.(\ref{eq8}), with c = 50 MeV, leads to a
suppression
of the cross section away from the data.

We have argued earlier that the dominance of the S$_{11}$
contributions is used to limit the forgoing comparisons and
discussions to this contribution. In fig 7 we revisit this
question. In part (a)  we show the results for the case of
production on C for the S$_{11}$. We compare this with the
results in the lower panel (b) based on calculations in
which the small contributions from the D$_{13}$, the nucleon pole
and the t-channel vector meson diagrams are included. These
contributions, being small, are calculated using the local
approximation (the non-local corrections in this case are
expected to be negligible). 
Each of the curves in panel (a) has a counterpart in panel (b): 
The solid curves represent the free calculations, the dashed curves 
represent calculations in which the mass reduction is affected by 
a value b = 0.014. The dash-dotted curves represent calculations in 
which we  make changes in both mass and width of the resonance. 
We assumed the width is reduced to 150 MeV (from 160 MeV) and in the 
meantime the mass is reduced as above.
It is seen from the figures that
the inclusion of the additional contributions, though it
modifies the cross section slightly, does not alter the
character of the comparison with the data, from those involving 
only the S$_{11}$ contributions.

The preceding comparisons were based on calculations using
the optical potential DW1 to describe the final state
interaction of the $\eta$ meson. The calculations of the
cross sections are sensitive to the optical potential used.
For example we show in Fig.8 a comparison of the cross
sections calculated using one of the optical potentials
derived by Oset {\it et al.}\cite{4OSE.PRC}. 
These calculations of the eta
optical potential depend on the magnitude of the real part of S$_{11}$
self energy in the medium, and we have used the case for which
the real part of the 
S$_{11}$ self-energy  is -50 MeV. The cross sections are seen
to lie below those calculated using the DW1 potential. It is
clear that this type of uncertainty should be kept in mind
before a final verdict is gleaned form the data as to 
the extent  of any  medium
modifications of the properties of the resonance.

\section{conclusion}
\label{conclusion}

A relativistic non-local model for the inclusive
photoproduction of $\eta$
mesons from complex nuclei is introduced. This model is
based on the dominance of the contributions due to the
formation of the S$_{11}$ resonance. The inclusion of
non-local effects is the main improvement over our earlier
work which was applied to the limited data available at the
time. Moreover, the non-local approach makes it possible to
include mass and width parameters of the resonance that may
depend on the nuclear density. 

Calculations were first carried out using the
free mass and width for the resonance (i.e. no medium
effect). Comparison of the results of these calculations
with the data available for C, Al and Cu, reproduce the over
all shape of the angular distributions as well as the total
cross section data up to 1.0 GeV. It is worth noting that
our results are in somewhat better agreement with the recent data of
LNS than with the older data.

The influence of possible medium modifications of the
properties of the resonance was also studied. Simple density
dependent forms  for the mass and width of the resonance
were used to test the effect these have on the calculated 
cross sections and their accord with the
data.  Comparison with the data showed that
reductions in the mass of S$_{11}$ on a scale similar to that suggested
for the nucleon leads to cross sections that disagree
strongly with the data.  The data can accommodate much
weaker changes in the mass and some broadening of the width.
Although one can claim some scattered instances of
improvement with these changes, it is not possible to claim
that the data show clear evidence of medium modifications of
resonance parameters. The lack of this evidence is due in
part to the occasional disagreement
between different measurements. There is also the inherent
weakness in our present knowledge of the final state
interaction of the eta meson with nuclei (represented here
by the optical potential), which adds to the uncertainty.  
In conclusion we would estimate
that if any medium modifications of the properties of the
S$_{11}$  resonance existed, they would likely be rather
small. This is in agreement with the general conclusions reached by the authors of refs.\cite{BKE, kinoshita, LM, MC}.
\section*{Acknowledgments}
We Would like to thank T. Yorita and T. Kinoshita for their 
helpful communications regarding their data. We also are grateful to  
E. Sumber, from Computing and Network Services at University of Alberta 
for parallelizing 
our computer code and to West Grid for allowing us to use their 
facilities to perform our computations. This work was supported in part by 
the Natural Sciences and
            Engineering Research Council of Canada.

\newpage
\section* {Table Caption}

\noindent TABLE 1. Strengths, reduced radii and diffuseness for the 
relativistic scalar and vector mean-field potentials, respectively.

\section* {Figure Captions}

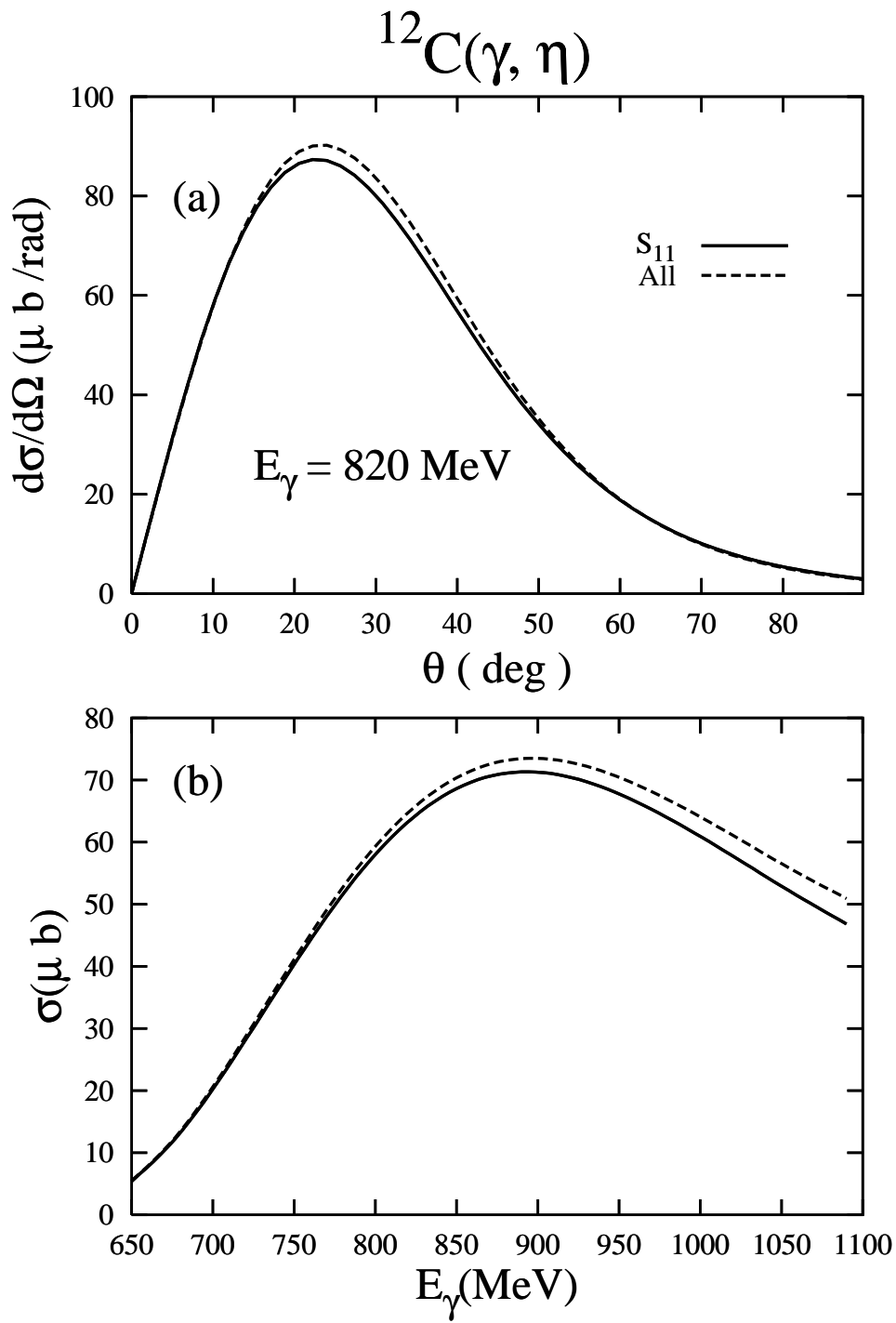
\noindent FIG. 1.  Differential (a) and total (b) inclusive cross section
for $^{12}C(\gamma, \eta)$ reaction. Solid curves: S$_{11}$ resonance 
contribution, dashed curves: Sum of contributions from 
S$_{11}$ resonance, D$_{13}$ resonance, 
nucleon pole and vector meson. 

\vspace{2.5 mm}
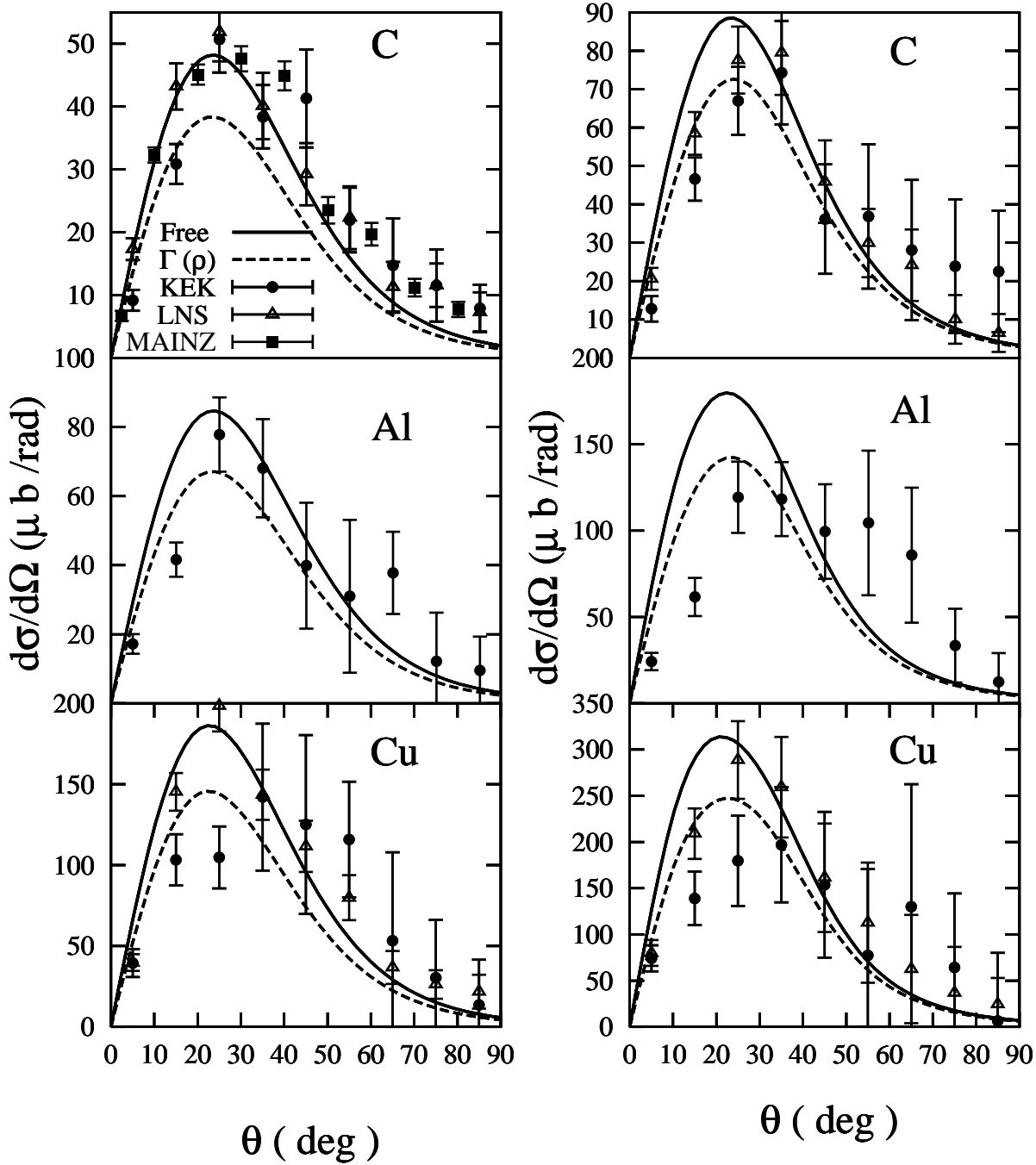
\noindent FIG. 2. Differential inclusive cross section
for $(\gamma, \eta)$ reaction on C, Al and Cu nuclei. 
The graphs in the left and right panels represent the 
results for incident photon energies of  
E$_\gamma$ = 740 MeV and  E$_\gamma$ = 980 MeV, respectively. 
Solid curves: calculations using the values 1535 MeV and 
160 MeV for the mass and width of S$_{11}$ 
resonance, respectively (labeled Free in this and subsequent figures), 
dashed curves: density dependent width of eq(\ref{eq8}) with c = 50 MeV 
 (labeled $\Gamma(\rho)$). 
Optical 
potential DW1 of  \protect ref.\cite{4LEE.NPA} is used. 
Data are 
(KEK) from \protect\cite{YORITA}, (LNS)  from \protect\cite{kinoshita} and 
(MAINZ) from \protect\cite{RLPLB}.

\vspace{2.5 mm}
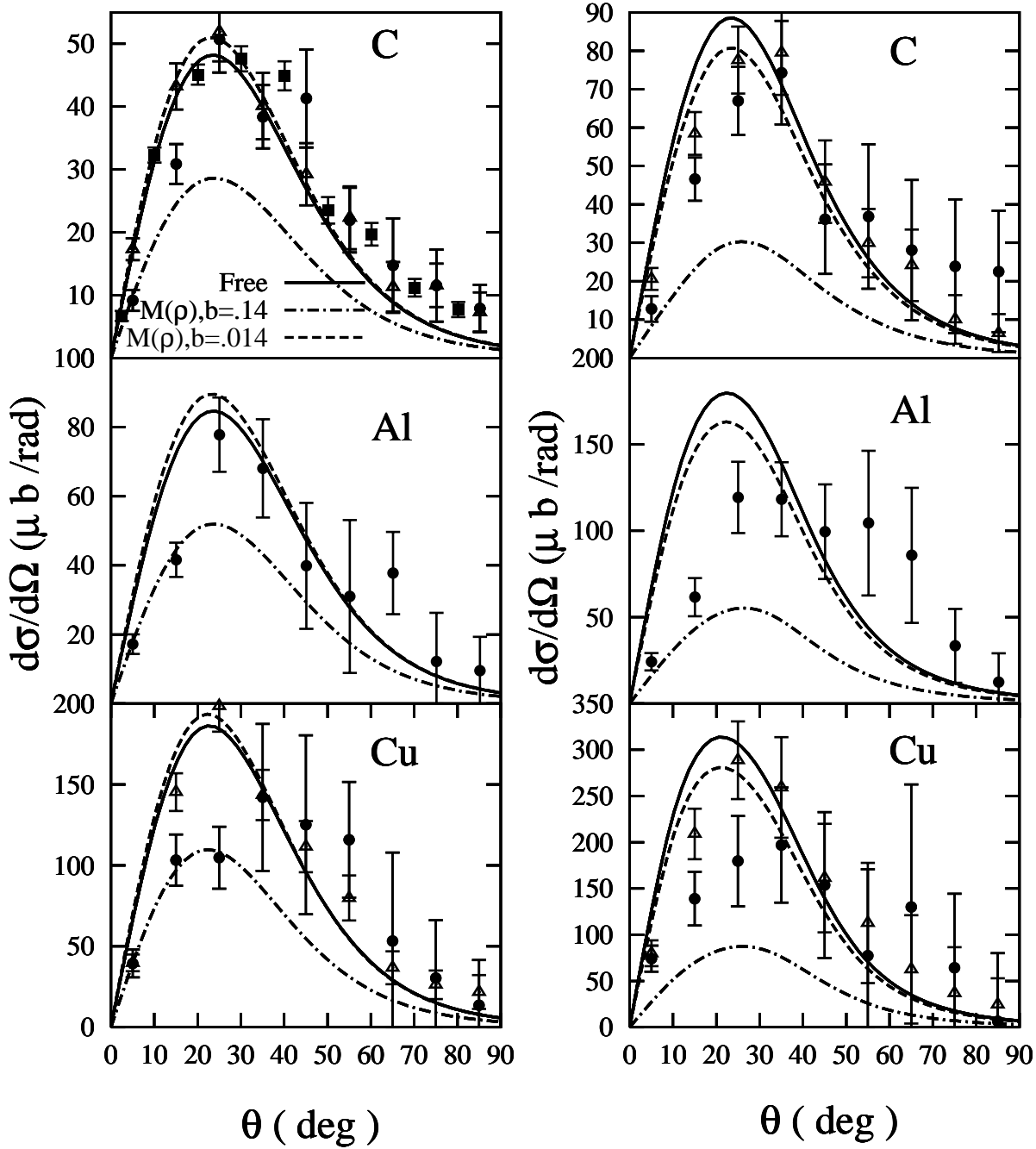
\noindent FIG. 3. Same as Fig.2, but for the effect of mass modifications.
 Solid curves: free calculations, 
dashed curves: density dependent mass of eq(\ref{eq7}) 
with b = 0.014, 
dash-dotted curves: density dependent mass of eq(\ref{eq7}) with b = 0.14.

\vspace{2.5 mm}
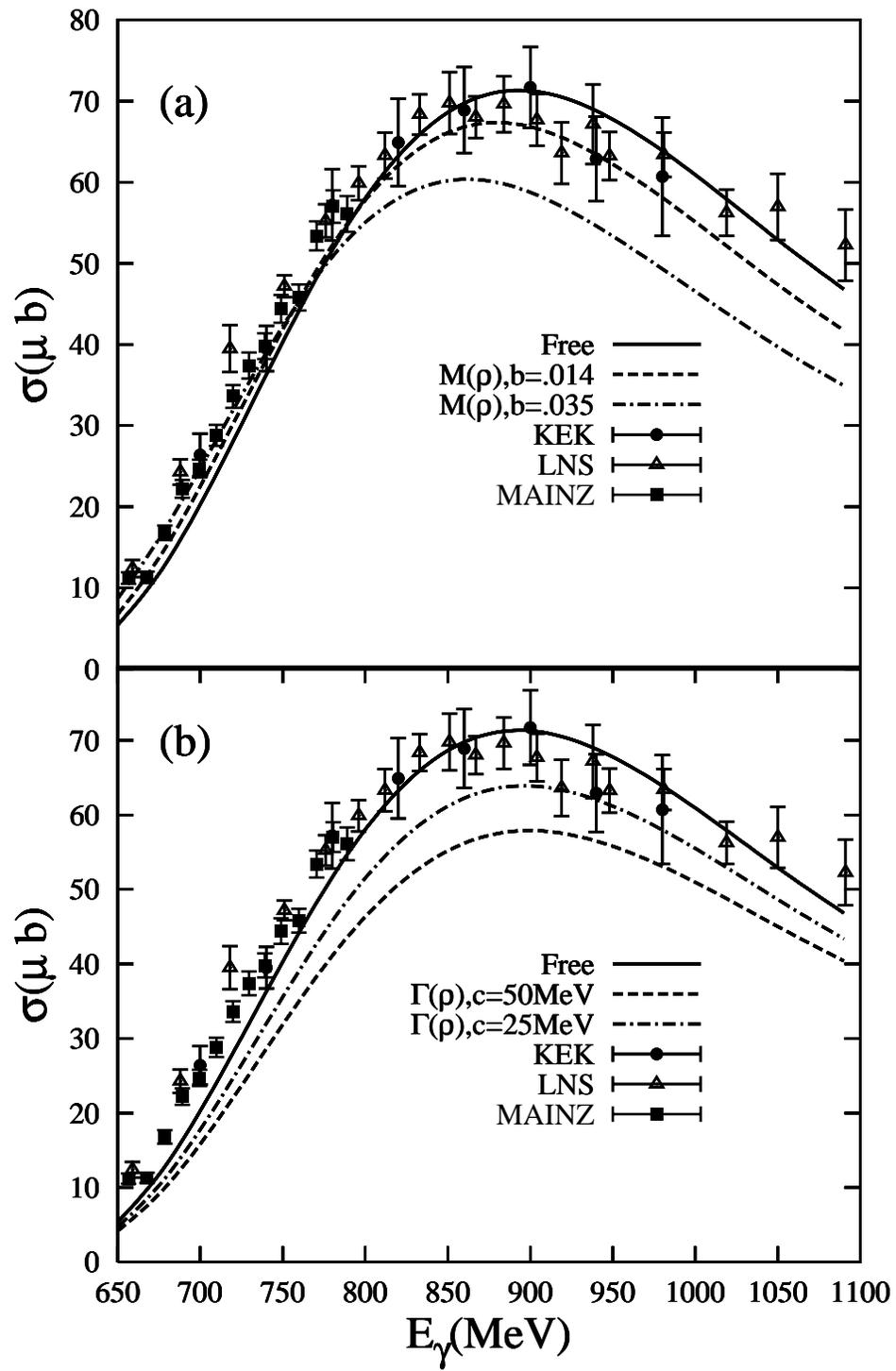
\noindent FIG. 4. Total inclusive cross section for the $^{12}C(\gamma, \eta)$ 
reaction. Data and potentials are from the 
same references as in Fig.2. Fig.4(a); solid curve:
free calculations, dashed curve: density dependent mass of eq(\ref{eq7}) 
with b=0.014,
dash-dotted curve: density dependent mass of eq(\ref{eq7}) with b=0.035.
Fig.4(b); solid curve: free calculation, dashed curve: density dependent 
width of eq(\ref{eq8}) 
with c = 50 MeV,
dash-dotted curve: density dependent width of eq(\ref{eq8}) with c = 25 MeV.

\vspace{2.5 mm}
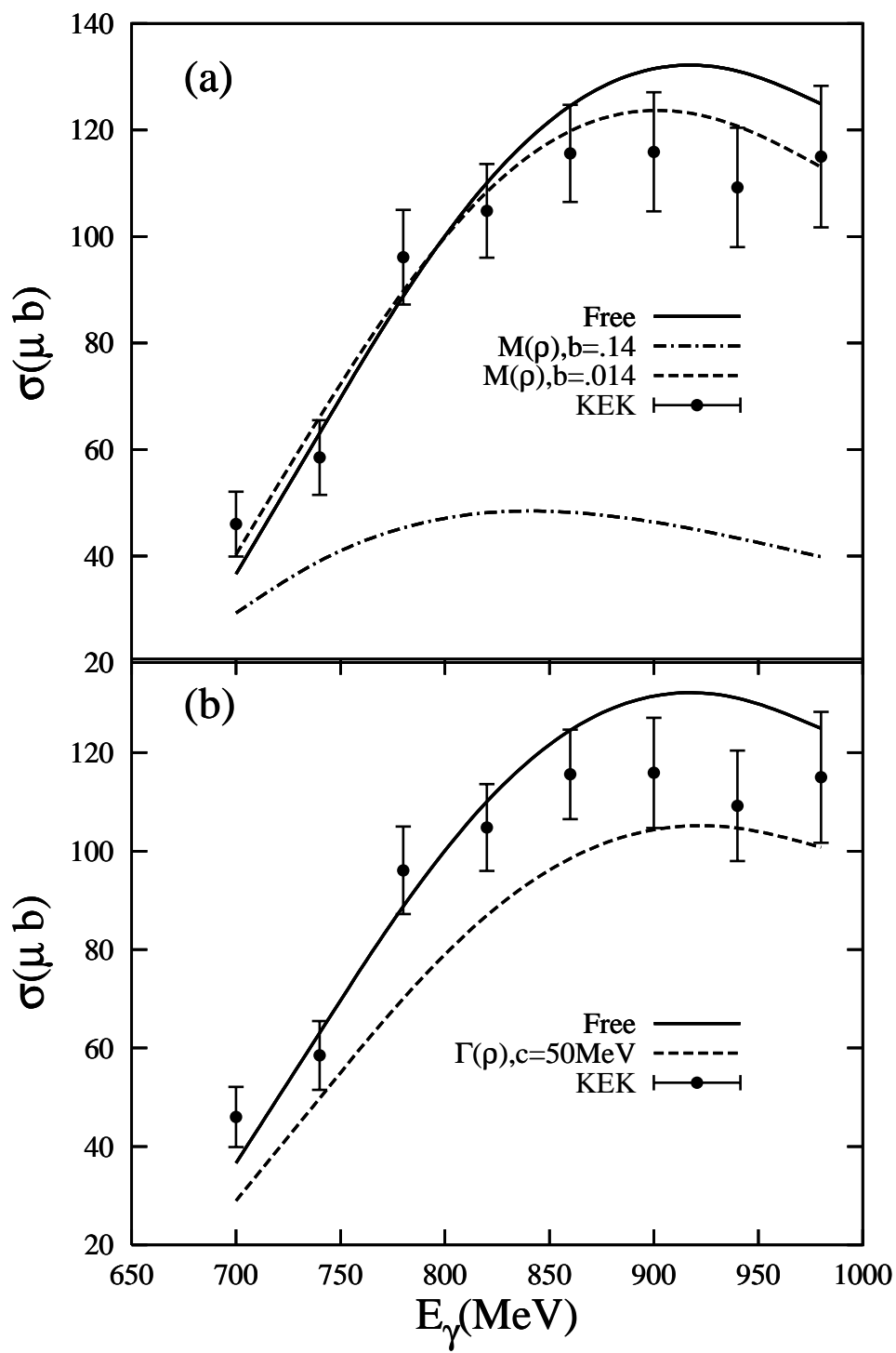
\noindent FIG. 5.  Same as Fig.4 but for Al. In Fig.5(a): 
solid curve: free calculations, dashed curve: 
density dependent mass of eq(\ref{eq7}) 
with b=0.014,
dash-dotted curve: density dependent mass of eq(\ref{eq7}) 
with b=0.14.
In Fig.5(b): solid curve: free calculations, dashed curve: 
density dependent width of eq(\ref{eq8}) 
with c = 50 MeV, data are those of ref.\protect\cite{YORITA}. 
 
\vspace{2.5 mm}
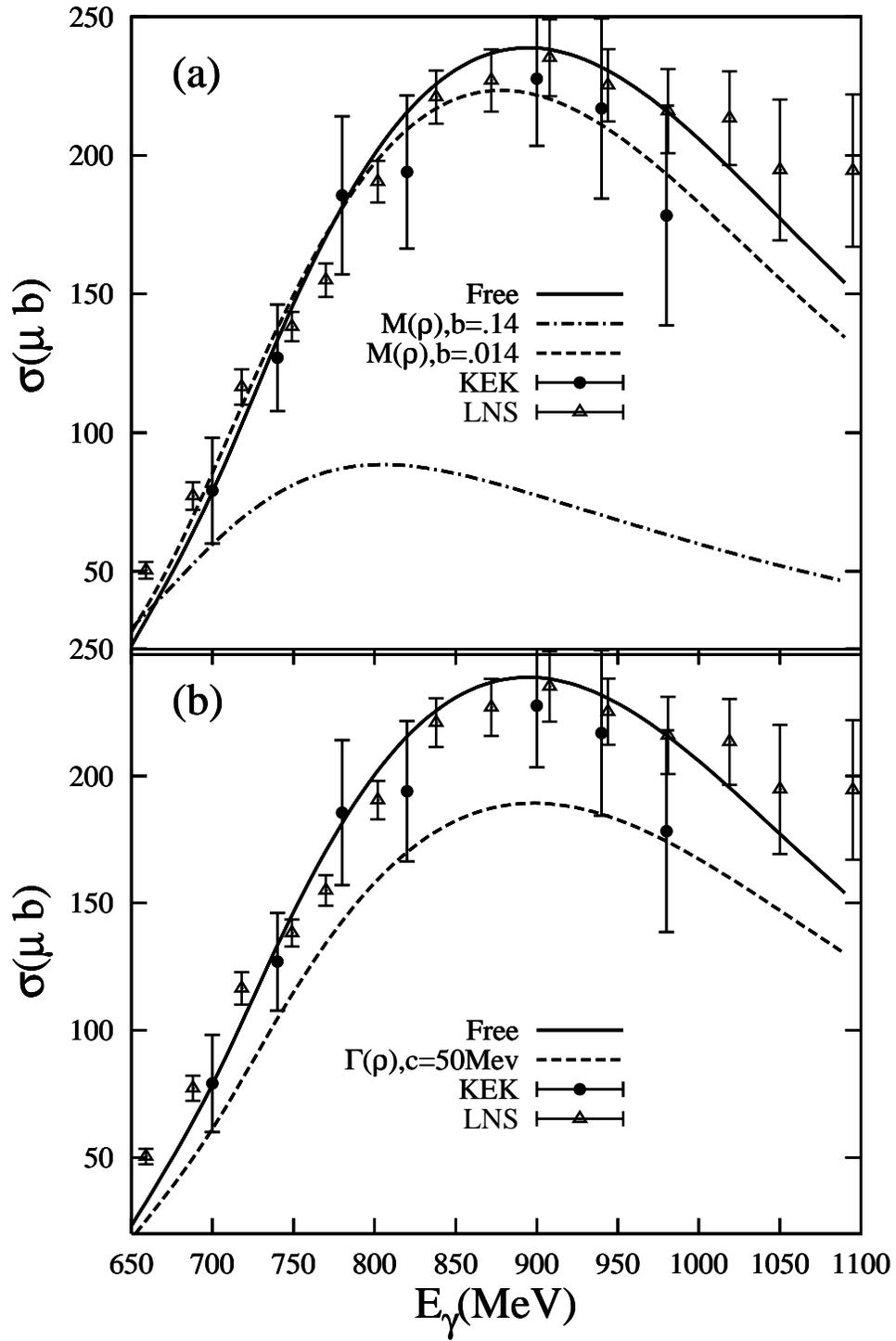
\noindent FIG. 6. Same as Fig.5 but for Cu. Curves are 
labeled same as Fig.5. Data are: (KEK) and (LNS) from refs.  
\protect\cite{YORITA} and   \protect\cite{kinoshita}, respectively. 

\vspace{2.5 mm}
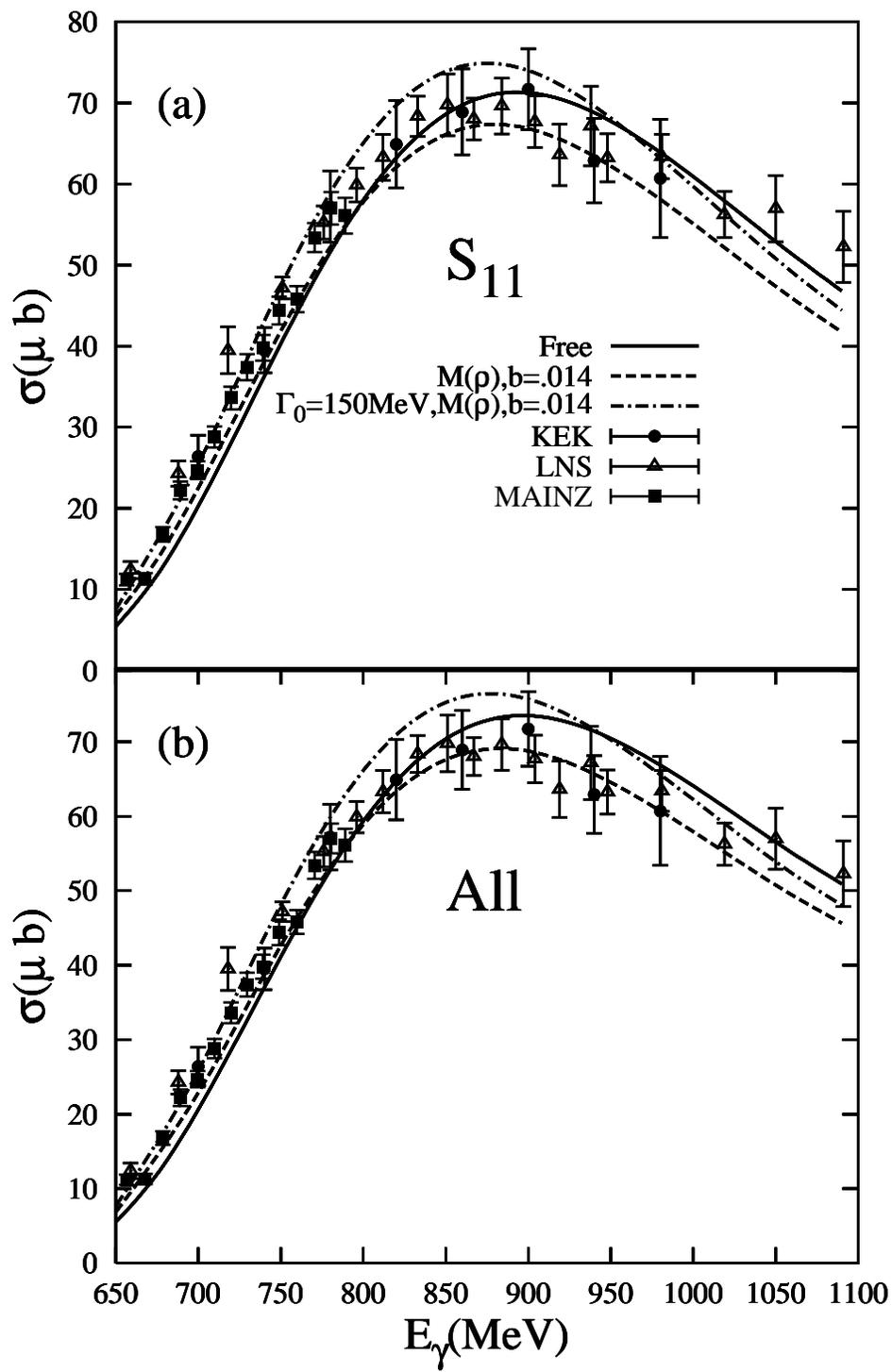
\noindent FIG. 7. Same as Fig.4.
 Curves in panel (a) are contributions from S$_{11}$ resonance, 
solid curve: free calculations,  
dashed curve: free width of 160 MeV and density 
dependent mass of eq(\ref{eq7}) 
with b=0.014,
dash-dotted curve: free width of 150 MeV and density dependent mass 
of eq(\ref{eq7}) with b=0.014.
The contributions of nucleon poles, D$_{13}$ resonance and 
vector meson diagrams are added to the ones for S$_{11}$ in the 
calculations presented in panel(b). Curves are labeled as in panel(a).

\vspace{2.5 mm}
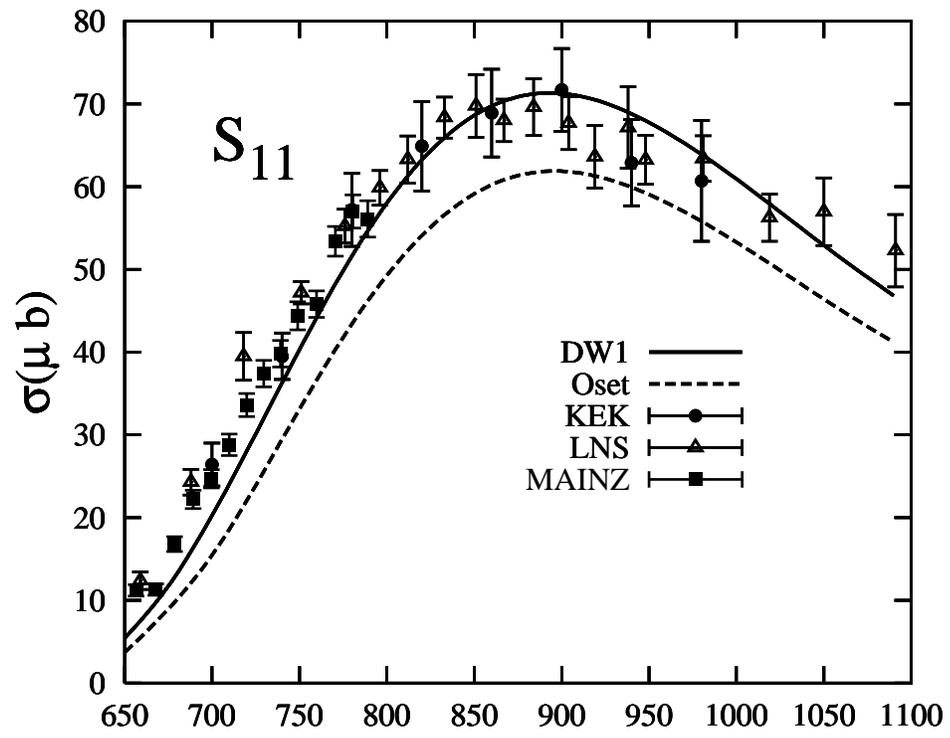
\noindent FIG. 8. Same as the free calculations in Fig.4 (S$_{11}$ resonance 
contributions only).
Solid curve: using optical 
potential DW1 of  \protect ref.\cite{4LEE.NPA}, dashed curve: 
using optical 
potential of  \protect ref.\cite{4OSE.PRC} with the real part of the S$_{11}$ 
self-energy set to -50 MeV.

\newpage

\begin{table}
\begin{center}
\begin{tabular}{c c c c c c c}\hline
Nucleus & V$_v$ &r$_v$& a$_v$&V$_s$&r$_s$&a$_s$  \\
        &MeV& (fm)&(fm)&(MeV)&(fm)&(fm)\\ \hline
$^{12}$C    & 385.7&1.056&0.427&-470.4&1.056&0.447           \\ \hline
$^{27}$Al    & 354.1& 1.09& 0.450&-444.5& 1.09& 0.500           \\ \hline
$^{63}$Cu    &348.1& 1.149& 0.476&-424.5& 1.149& 0.506           \\ \hline
\end{tabular}
\end{center}
\caption{}
\label{table1}
\end{table}

\end{document}